\documentclass[aps, prd, showpacs, superscriptaddress, preprintnumbers, nofootinbib, twocolumn]{revtex4-1}
\usepackage[pdftex]{graphicx, color}
\usepackage{amsmath, amssymb, amscd, latexsym, bm, braket}
\usepackage[colorlinks=true,pdfstartview=FitV,linkcolor=blue,citecolor=magenta,urlcolor=blue,bookmarks=true,bookmarksnumbered=true]{hyperref}

\begin{document}
\title{Mutual assistance between the Schwinger mechanism and the dynamical Casimir effect}

\author{Hidetoshi Taya}
\email{h\_taya@keio.jp}
\address{Department of Physics and Center for Field Theory and Particle Physics, Fudan University, Shanghai, 200433, China }
\address{Research and Education Center for Natural Sciences, Keio University 4-1-1 Hiyoshi, Kohoku-ku, Yokohama, Kanagawa 223-8521, Japan}

\date{\today}

\begin{abstract}
We study massive charged particle production from the vacuum confined between two vibrating plates in the presence of a strong electric field.  We analytically derive a formula for the production number based on the perturbation theory in the Furry picture, and show that the Schwinger mechanism by the strong electric field and the dynamical Casimir effect by the vibration assist with each other to dramatically enhance the production number by orders of the magnitude.  
\end{abstract}


\maketitle

\section{Introduction}

According to the quantum field theory, our vacuum is by no means a simple empty space, but should be regarded as a sort of matter, in which virtual particles are ceaselessly created and annihilated.  Our vacuum, therefore, exhibits non-trivial responses when exposed to external forces/fields just as ordinary matters do.  In particular, if the external forces/fields are strong, the virtual particles in the vacuum may acquire large energy through the interaction with the forces/fields.  If the acquired energy is sufficiently large, i.e., larger than the mass scale of the particles, the virtual particles become on-shell real particles.  Therefore, our vacuum is no longer a stable state in the presence of strong forces/fields and decays spontaneously by producing particles.

An example of such a particle production mechanism is the Schwinger mechanism, which is driven by a strong slow electric field (for review, see, e.g., Refs.~\cite{dun05, ruf10, gel15}).  The original idea of the Schwinger mechanism was first proposed by Sauter in 1931 \cite{sau31} as a resolution to the Klein paradox in the barrier scattering problem of a Dirac particle \cite{kle29}.  The idea was that a strong slow electric field induces a level crossing between positive and negative energy states.  The level crossing, then, causes quantum tunneling from a negative energy state to a positive one, which can be interpreted as pair production of a particle and an anti-particle.  This production process is genuinely non-perturbative since it is driven by quantum tunneling unless the electric field is fast \cite{bre70, pop72, tay14}.  Sauter's idea was, then, sophisticated by Heisenberg and Euler by deriving the one-loop effective action in the presence of a constant and homogeneous strong electromagnetic field \cite{hei36} and by Schwinger by fully formulating Sauter's idea and Heisenberg and Euler's calculation within the quantum field theory \cite{sch51}.  Later, the Schwinger mechanism for Dirac particles was generalized to scalar particles \cite{wei36}, vector particles \cite{van65}, and particles with arbitrary spin \cite{mar72}, which revealed that the Schwinger mechanism is insensitive to spin or boson/fermion statistics provided that there is no magnetic field and that the electric field does not change its direction.  Despite its long history, the Schwinger mechanism is still a hot research topic in theoretical physics and has been developing continuously.  To name a few, examples include the following: study of finite size effects \cite{wan88, gav16, ceh95, pav91, sch90, cec88, cec89, gat92, war92}; inclusion of back-reaction effects \cite{mat85, gat87, coo89, klu91, klu92, klu93, blo99, rau94, smo97}; proposal of avalanche-like particle production mechanism (QED cascade) \cite{bel08, fed10, bul11, ner10, elk11}; developments of new theoretical techniques (e.g., the world-line instanton method \cite{dun05a, dun06a, dun06b}, classical statistical method \cite{gel13, kas14}, and the perturbation theory in the Furry picture \cite{tay19, hua19, tay20, gre17, gre18, gre19}); and phenomenological applications (e.g., the early stage dynamics of ultra-relativistic heavy-ion collisions \cite{gel06, tan09, tan11, tay17, gle83, gat87, cas79, cas80, mat85, rug15b, rug15a}, magnetogenesis in the early Universe \cite{kob14, sob18, sha17, sor19, sta18, kob19}, and spin-current generation in spintronics \cite{hua19a}).

From an experimental side, the Schwinger mechanism has not been observed yet.  Since the lightest charged particle is electron, it requires extremely strong electric field $eE_{\rm cr} \equiv m_{\rm e}^2 \sim \sqrt{10^{29}\;{\rm W/cm^2} }$ for the Schwinger mechanism to be manifest.  The available electric field strength in the current laboratory experiments is, however, far below this critical value.  In fact, $eE \sim \sqrt{10^{22}\;{\rm W/cm^2}}$ is the present world record established by the highest-intensity focused laser at HERCULES \cite{yan08}.  The forthcoming laser experiments at, e.g., the Extreme Light Infrastructure (ELI) and at the Exawatt Center for Extreme Light Studies (XCELS), may reach $eE \sim \sqrt{ 10^{24{\rm -}25}\;{\rm W/cm^2}}$, which is still below the critical field strength $eE_{\rm cr}$ by several orders of the magnitude.  Therefore, much attention has been paid recently to how to enhance the Schwinger mechanism to make it experimentally observable.  One of the most promising ideas is the so-called dynamically assisted Schwinger mechanism \cite{sch08, piz09, dun09, mon10a, mon10b} (see also Refs.~\cite{koh13, ott15, vil19}), which proposed that a superposition of a fast weak electromagnetic field can enhance the Schwinger mechanism by several orders of the magnitude.  Recently, Refs.~\cite{tay19, hua19, tay20, gre17, gre18, gre19} showed that the enhancement can be captured very well within the perturbation theory in the Furry picture \cite{fur51, fra81, fra91} and clarified that the parametric (or perturbative) particle production by the superimposed field is the essence of the enhancement.  This observation implies that a similar enhancement mechanism may occur not only with a fast weak electromagnetic field but also with other kinds of forces/fields involving parametric particle production.

Another example of the particle production from the vacuum is the dynamical Casimir effect (for review, see, e.g., Refs.~\cite{dod01, dod10}), which is driven by vibrating plates (or, more generally, variations of geometries in time, e.g., acceleration of a single mirror \cite{ful76} and rotation of an object \cite{zel71, mag12}).  The dynamical Casimir effect, named by Yablonovitch \cite{yab89} and Schwinger \cite{sch92}, was first proposed by Moore in 1970 \cite{moo70}.  It is a parametric effect, which occurs when the typical frequency of the vibration matches the mass gap (or the eigen-frequency of cavity modes inside the boundaries) and can essentially be studied with perturbative approaches as long as the amplitude of the vibration is sufficiently small \cite{for82, dod93, dod96, mun98}.  Notice that these features are quite distinct from the Schwinger mechanism, which is not a parametric effect and hence does not exhibit any threshold behaviors.  Also, a non-perturbative treatment of a strong electric field is essential to describe the Schwinger mechanism.  The original idea of the dynamical Casimir effect proposed by Moore was for massless photon production, but was naturally generalized to massive particle production, e.g., in Refs.~\cite{raz83, reu89, dit02, fri13, dou15}.

The direct experimental observation of the dynamical Casimir effect is still lacking, which requires a very fast vibration of plates, e.g., $\omega/2\pi \sim 10\;{\rm GHz}$ to produce cavity photons (which could be lowered by non-perturbative effects \cite{mac18}) and much larger frequency for massive particles like electron $\omega/2\pi \sim 2m_e/2\pi \sim 100\;{\rm EHz}$.  Such a fast vibration is not experimentally available at the present, although a close value $\omega/2\pi \sim 6\;{\rm GHz}$ was achieved recently in Ref.~\cite{con10}.  On the other hand, there already exists an {\it indirect} observation of the dynamical Casimir effect utilizing some effective motion instead of real vibration of plates.  For example, Ref.~\cite{wil11} modulated {\it electrical} length of a coplanar transmission line by modulating inductance of a superconducting quantum interference device (SQUID) connected to the transmission line \cite{joh09, joh10}.  This effectively realized $\omega/2\pi \sim 11\;{\rm GHz}$-vibration, which enabled us to observe the cavity photon production by the dynamical Casimir effect for the first time.  It is, however, still very difficult even with the indirect methods to observe the dynamical Casimir effect for massive particles.  Hence, a new mechanism to lower the threshold frequency is highly demanded for future experiments.

The purpose of this paper is to propose a new mechanism to enhance the Schwinger mechanism and the dynamical Casimir effect for a massive charged particle.  Namely, we consider a charged massive scalar field confined inside of two vibrating plates and apply a strong external electric field.  With this setup, we show that the dynamical Casimir effect by the vibrating plates assists the Schwinger mechanism by the strong electric field and vice versa.  In particular, we show that the assistance effectively reduces the mass gap, and the particle production is dramatically enhanced by orders of the magnitude.  This may be understood as an analog of the dynamically assisted Schwinger mechanism \cite{sch08, piz09, dun09, mon10a, mon10b}, in which the combination of the parametric particle production mechanism and the non-perturbative Schwinger mechanism enhances the particle production \cite{tay19, hua19, tay20, gre17, gre18, gre19}.  Our theory is based on the perturbation theory in the Furry picture \cite{gre18, fra91, fra81, fur51, gre19, gre17, tay19, tay20, hua19}, in which the interaction with the electric field is treated non-perturbatively but that with the vibration perturbatively.  The perturbation theory in the Furry picture is very successful in the analytical description of the dynamically assisted Schwinger mechanism \cite{tay19, hua19, tay20, gre17, gre18, gre19} and, as we shall show in this paper, can smoothly describe the interplay between the non-perturbative Schwinger mechanism and the parametric dynamical Casimir effect, which is not feasible within adiabatic approaches such as the worldline instanton method \cite{dun05a, dun06a, dun06b}.  Note that our formalism can naturally be extended to, e.g., fermionic case and particles with arbitrary spin, which would be discussed in another publication.

This paper is organized as follows: In Sec.~\ref{sec2}, we derive a general formula for the particle production number in the presence of both vibrating plates and a strong electric field based on the perturbation theory in the Furry picture.  In Sec.~\ref{sec3}, we discuss quantitative features of the production number by considering a constant and homogeneous electric field configuration, for which the general formula can be evaluated exactly.  We show that the dynamical Casimir effect by the vibrating plates assists the Schwinger mechanism by the strong electric field and vice versa, which results in an enhancement of the production number by orders of the magnitude.  Section~\ref{sec4} is devoted to summary and discussion.

\section{General formula} \label{sec2}

We derive a production number formula for a massive charged scalar field in the presence of both vibrating plates and a strong electric field based on the perturbation theory in the Furry picture \cite{gre18, fra91, fra81, fur51, gre19, gre17, tay19, hua19, tay20}.  We first explain our physical setup in Sec.~\ref{sec2a}, and write down the corresponding field equation in Sec.~\ref{sec2b}.  In Sec.~\ref{sec2C}, we solve the field equation perturbatively in terms of the vibration while the non-perturbative interaction with the strong electric field is fully taken into account by using a Green function dressed by the strong electric field (the perturbation theory in the Furry picture).  Then, we canonically quantize the field operator at asymptotic states, which allows us to directly compute the vacuum expectation value of the number operator to derive the production number formula.

\subsection{Setup} \label{sec2a}

\begin{figure}[!t]
\begin{center}
\includegraphics[trim={1 50 60 20}, clip, width=0.49\textwidth]{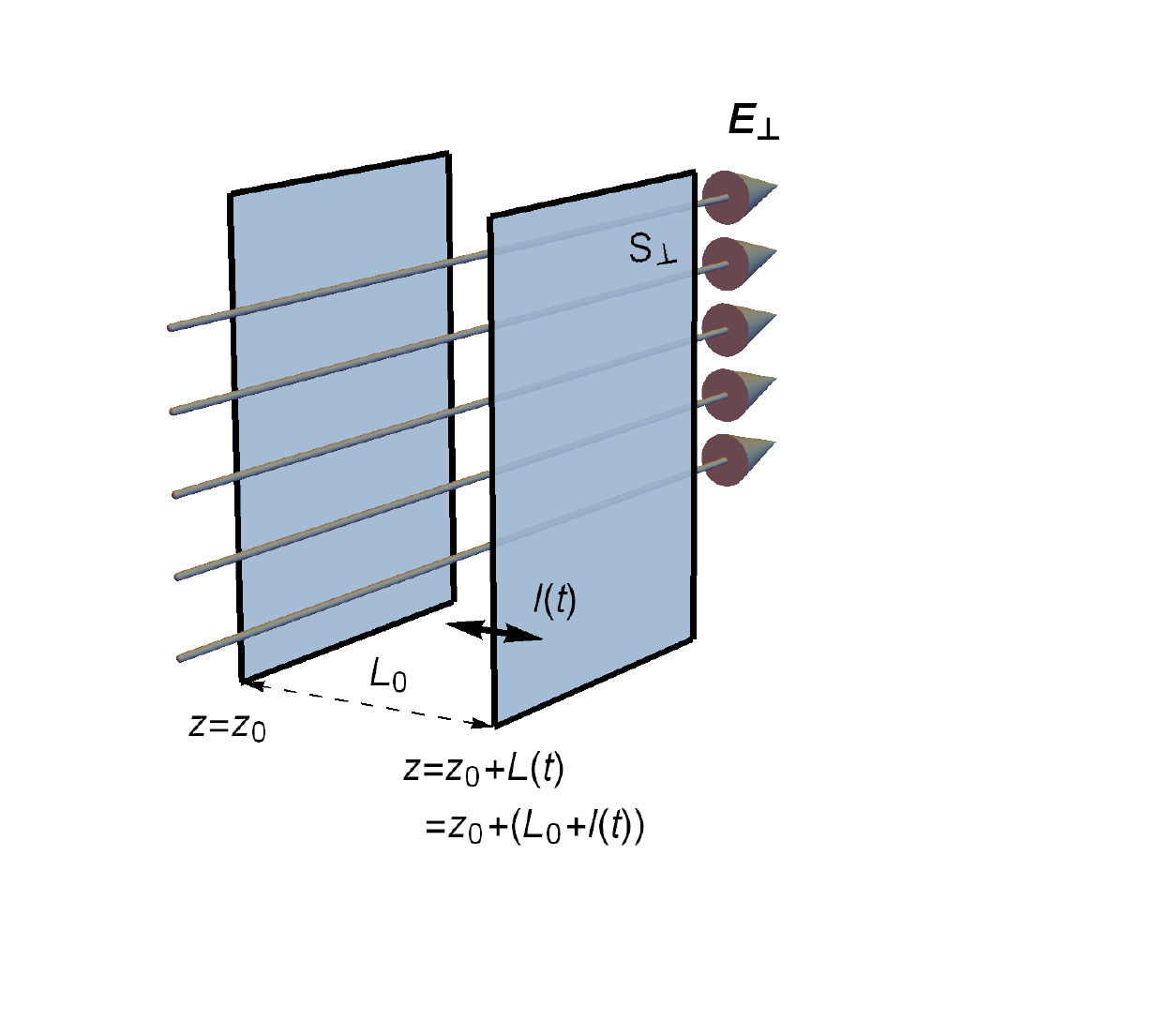}
\caption{\label{fig1} A schematic picture of our system: a strong electric field ${\bm E}_{\perp}$ applied between two vibrating plates with transverse area $S_{\perp}$ and distance $L(t)=L_0+l(t)$.  }
\end{center}
\end{figure}

We consider a massive charged scalar field $\hat{\phi}$ with mass $m$ and electric charge $e$ confined between two vibrating plates with (infinitely large) transverse area $S_\perp$ put at $z=z_0, z_0+L(t)$, and apply a strong electric field in the transverse direction (see Fig.~\ref{fig1}), 
\begin{align}
	A^{\mu} \equiv (0, {\bm A}_{\perp}, 0) = (0, -\int^t {\rm d}t\;{\bm E}_{\perp}, 0),  
\end{align}
where we adopted the temporal gauge $A^0=0$.  One can always gauge out the longitudinal component $A^3 = 0$ because there is no longitudinal electric field.

We assume (i) that the plates are vibrating just slightly, i.e., the amplitude of the vibration is small (but the frequency can be large); and (ii) that the plates adiabatically stop moving at the infinite future and past.  Namely, we separate $L(t)$ into a time-independent part $L_0$ and a time-dependent part $l(t)$ as 
\begin{align}
	L(t) \equiv L_0 + l(t), \label{eq--2}
\end{align}
and assume (i) smallness of the amplitude, 
\begin{align}
	\left| \frac{l}{L_0} \right| \ll 1,    
\end{align}
and (ii) adiabaticity,
\begin{align}
	\lim_{ t \to \pm \infty} l = 0.  \label{eq-4}
\end{align}
These assumptions may be reasonable because (i) it is experimentally difficult to realize a vibration with large amplitude if its frequency is large, and (ii) it is impossible to vibrate the plates forever.  Theoretically, as we describe below, we use (i) to justify the lowest order perturbation in terms of the vibration $l$; and (ii) to define asymptotic states in a well-defined manner.  Note that Eq.~(\ref{eq--2}) does not necessarily mean $L_0$ is large, i.e., only the ratio $l/L_0$ is assumed to be small and the size of $L_0$ is arbitrary.

\subsection{Field equation and boundary condition} \label{sec2b}

The field operator $\hat{\phi}$ obeys the Klein-Gordon equation,
\begin{align}
	0 = \left[  \partial_t^2 - ({\bm \partial}_{\perp} - {\rm i} e {\bm A}_{\perp})^2 - \partial_z^2  + m^2 \right] \hat{\phi}.  \label{eq1}
\end{align}
As the field $\hat{\phi}$ cannot go outside of the two plates, $\hat{\phi}$ should satisfy a boundary condition 
\begin{align}
	0 = \hat{\phi}(z_0) = \hat{\phi}(z_0 + L(t)).   \label{eq2}
\end{align}

The time-depending boundary condition (\ref{eq2}) is quite inconvenient in solving the Klein-Gordon equation (\ref{eq1}), and to see how the vibrating plates affect the time-evolution of the system.  To circumvent these inconveniences, we change the coordinates as
\begin{align}
	\begin{pmatrix} t \\ z \end{pmatrix}
	\to 
	\begin{pmatrix} \tau(t,z) \\ \xi(t,z) \end{pmatrix} 
	\equiv 
	\begin{pmatrix} t \\ (z - z_0)/L(t) \end{pmatrix} .
\end{align}
Then, the Klein-Gordon equation (\ref{eq1}) and the boundary condition (\ref{eq2}) can be re-expressed as
\begin{align}
	0 &= \Biggl[ \left(  \frac{\partial}{\partial \tau} - \frac{\dot{L}}{L} \xi  \frac{\partial}{\partial \xi}  \right)^2  \nonumber\\
	  &\quad\quad  -  ({\bm \partial}_{\perp} - {\rm i}e{\bm A}_{\perp})^2 -  \left(   \frac{1}{L} \frac{\partial}{\partial \xi} \right)^2 +  m^2 \Biggl] \hat{\phi} \label{eq7}
\end{align}
and
\begin{align}
	0 = \hat{\phi}(\xi = 0) = \hat{\phi}(\xi = 1).  \label{eq8}
\end{align}
Notice that $L$ explicitly enters in the new field equation (\ref{eq7}), which effectively describes the interaction between particles and the vibrating plates.  Through the interaction, the vibrating plates can supply energy to the vacuum if $\dot{L} \neq 0$, and thus particles can be excited from the vacuum if the supplied energy is larger than the mass gap $\sim m$ (i.e., the dynamical Casimir effect).  Mathematically speaking, the time-translational invariance in Eq.~(\ref{eq7}) is explicitly broken by $L$, so that the positive/negative frequency modes (particle and anti-particle modes) $\propto {\rm e}^{\mp {\rm i}\omega_{\bm p}t}$ are mixed with each other during the time-evolution.  As shown below, the vacuum expectation value of the number operator can have non-zero value because of this mixing.

\subsection{Production number within the perturbation theory in the Furry picture} \label{sec2C}

We evaluate the production number of particles and anti-particles from the vacuum in the presence of the vibration and strong electric field: We first solve the field equation (\ref{eq7}) within the perturbation theory in the Furry picture (i.e., a perturbative calculation with a dressed propagator/wavefunction) \cite{gre18, fra91, fra81, fur51, gre19, gre17, tay19, hua19, tay20}.  As we assumed that the plates are vibrating just slightly, we are justified to expand the field $\hat{\phi}$ perturbatively in terms of the small displacement $l$.  On the other hand, the electric field ${\bm E}_{\perp}$ is assumed to be strong, so that its interaction with the field $\hat{\phi}$ should be treated non-perturbatively.  That is, a free propagator/wavefunction is inappropriate to perform the perturbative calculation in terms of the vibration, and one has to use a dressed propagator/wavefunction which is fully dressed by the strong electric field.  After obtaining a solution of the field equation (\ref{eq7}) within the perturbation theory in the Furry picture, we canonically quantize the field operator to define annihilation operators at the asymptotic future and past.  Then, we can explicitly evaluate the in-vacuum expectation value of the number operator at $t = \infty$ to get a production number formula up to the lowest order in the displacement $l$, while the interaction with the strong electric field is fully taken into account.

We solve the field equation (\ref{eq7}) within the perturbation theory in the Furry picture.  To this end, we first expand Eq.~(\ref{eq7}) in terms of the small displacement $l \ll L_0$.  We find
\begin{align}
	&\left[ \frac{\partial^2}{\partial \tau^2}  -  ({\bm \partial}_{\perp} - {\rm i}e{\bm A}_{\perp})^2  - \left( \frac{1}{L_0}\frac{\partial}{\partial \xi} \right)^2 + m^2 \right] \hat{\phi}  = V \hat{\phi}, \label{eq15}  
\end{align}
where 
\begin{align}
	V &\equiv \left(  \frac{\ddot{l}}{L_0} + 2 \frac{\dot{l}}{L_0}\frac{\partial }{\partial \tau} \right) \xi \frac{\partial}{\partial \xi} \nonumber\\
	  &\quad- 2 \frac{l}{L_0} \left( \frac{1}{L_0} \frac{\partial}{\partial \xi} \right)^2 + {\mathcal O} \left( \left(l/L_0 \right)^2 \right) \label{eq16}
\end{align}
describes effects of the vibration.  Note that $L_0^{-1}\partial_\xi$ gives the longitudinal momentum and is totally independent of the smallness of $l$.  Also, note that $V$ should be adiabatically switched off at the infinite future and past,
\begin{align}
	\lim_{ \tau \to \pm \infty} V = 0, \label{eq23}
\end{align}
because of the adiabaticity of $l$ (\ref{eq-4}).  Then, by using the Green function technique, one can write down a formal solution of Eq.~(\ref{eq15}) as
\begin{align}
	\hat{\phi}(x)	&= \sqrt{Z} \hat{\phi}^{\rm in}(x) + \int {\rm d}^4y\; G_{\rm R}(x,y) V(y) \hat{\phi}(y) \nonumber\\
			&= \sqrt{Z}\Bigl[  \hat{\phi}^{\rm in}(x) + \int {\rm d}^4 y\; G_{\rm R}(x,y) V(y) \hat{\phi}^{\rm in} (y) \nonumber\\
			&\quad\quad\quad + {\mathcal O} \left( \left(l/L_0 \right)^2 \right) \Bigl], \label{eq20}
\end{align}
where $Z$ is the field renormalization constant and $G_{\rm R}$ is the retarded Green function satisfying
\begin{align}
	\left\{ \begin{array}{l}
		\delta^4(x-x') = \left[ \frac{\partial^2}{\partial \tau^2}  -  ({\bm \partial}_{\perp} - {\rm i}e{\bm A}_{\perp})^2 \right. \\
		\quad\quad\quad\quad\quad\quad \left.  - \left( \frac{1}{L_0}\frac{\partial}{\partial \xi}  \right)^2 + m^2 \right] G_{\rm R}(x,x')  \\
		0 = G_{\rm R}(\tau<\tau') 
	\end{array} \right. .
\end{align}
Notice that the Green function $G_{\rm R}$ is fully dressed by the strong electric field ${\bm E}_{\perp}$.  In Eq.~(\ref{eq20}), we required the Lehmann-Symanzik-Zimmermann (LSZ) asymptotic condition \cite{leh55} onto $\hat{\phi}$ as\footnote{Strictly speaking, the equality in Eq.~(\ref{eq24}) should be interpreted as a {\it weak} equality: It is not a strong equality among operators, but a weak equality holds only after operators are sandwiched by states.  In this work, we are interested only in expectation values of the number operator, so that the difference between a strong and weak equality is not important.  }
\begin{align}
	\lim_{ \tau \to \pm \infty} \left[ \hat{\phi}(x) - \sqrt{Z} \hat{\phi}^{\rm as} (x)   \right] = 0, \label{eq24}
\end{align}
where $\hat{\phi}^{\rm as}$ (${\rm as} = {\rm in/out}$) is a solution of the field equation without $V$ (i.e., without the perturbative vibration) because $V$ is vanishing at the infinite future and past [see Eq.~(\ref{eq23})].

In order to define the notion of a particle from the solution of the field equation (\ref{eq20}), we express the asymptotic field operator $\hat{\phi}^{\rm as}$ in terms of a mode integral as
\begin{align}
	\hat{\phi}^{\rm as}(x) 
	&= \frac{2\pi}{L_0}\sum_{n=1}^{\infty} \int {\rm d}^2{\bm p}_{\perp} \nonumber\\
	&\quad\times \Bigl[   {}_+ \phi^{\rm as}_{{\bm p}_{\perp},n}(x) \hat{a}^{\rm as}_{{\bm p}_{\perp},n}  +   {}_- \phi^{\rm as}_{{\bm p}_{\perp},n}(x) \hat{b}^{{\rm as}\dagger}_{-{\bm p}_{\perp},n} \bigl],   
\end{align}
where ${}_{\pm} \phi^{\rm as}_{{\bm p}_{\perp},n}$ are the positive and negative frequency mode functions at the corresponding asymptotic times which are defined as two independent solutions of 
\begin{align}
	\left[ \frac{\partial^2}{\partial \tau^2}  -  ({\bm \partial}_{\perp} - {\rm i}e{\bm A}_{\perp})^2 - \frac{1}{L_0^2}\frac{\partial^2}{\partial \xi^2} + m^2 \right] {}_{\pm} \phi^{\rm as}_{{\bm p}_{\perp},n} = 0  , \label{eq25}
\end{align}
with a boundary condition (i.e., a plane wave at the infinite future and past),
\begin{align}
	\lim_{\tau \to -\infty/+\infty } {}_{\pm} \phi^{\rm in/out}_{{\bm p}_{\perp},n}			&=\frac{ \exp \left[ \mp {\rm i} \int^{\tau} {\rm d}\tau\; \omega_{{\bm p}_{\perp}-e{\bm A}_{\perp},n} \right] }{\sqrt{2\omega_{{\bm p}_{\perp}-e{\bm A}_{\perp},n}}} \nonumber\\
	&\quad \times \frac{{\rm e}^{i{\bm p}_{\perp}\cdot{\bm x}_{\perp}}}{2\pi} \sqrt{\frac{L_0}{\pi}} \sin (n\pi \xi) , 
\end{align}
where $\omega_{{\bm p}_{\perp},n}$ ($n=1,2,\ldots$) is the on-shell energy, 
\begin{align}
	\omega_{{\bm p}_{\perp},n} \equiv \sqrt{ m^2 + {\bm p}_{\perp}^2 + \left(\frac{n\pi}{L_0} \right)^2 }.  \label{eq-19}
\end{align}
Here, we can safely identify the positive and negative frequency mode functions at the asymptotic times by the plane waves thanks to the adiabatic assumption (\ref{eq23}): The vibration adiabatically goes off, and the time-translational symmetry is restored at the asymptotic times.  Therefore, the plane waves, which are the eigenfunctions of the time-translational operation $-{\rm i}\partial_{\tau}$, becomes a good basis for the mode expansion at the asymptotic times.  It should be stressed here that ${}_{\pm} \phi^{\rm in}_{{\bm p}_{\perp},n} \neq {}_{\pm} \phi^{\rm out}_{{\bm p}_{\perp},n}$ if $L$ or ${\bm A}_{\perp}$ is time-dependent because their time-dependence mixes the positive and negative frequency modes.  Only if $L$ and ${\bm A}_{\perp}$ are time-independent, can one have ${}_{\pm} \phi^{\rm in}_{{\bm p}_{\perp},n} = {}_{\pm} \phi^{\rm out}_{{\bm p}_{\perp},n}$.  Note that the plane waves contain a sine function and the on-shell energy is quantized by the label $n=1,2,\cdots \in {\mathbb N}$ because of the boundary condition (\ref{eq8}).  We also note that the mode functions ${}_{\pm} \phi^{\rm as}_{{\bm p}_{\perp},n}$ are normalized as
\begin{align}
\left\{ \begin{array}{l}
	\displaystyle ( {}_{\pm}\phi_{{\bm p}_{\perp},n}^{{\rm as}} | {}_{\pm}\phi_{{\bm p}'_{\perp},n'}^{{\rm as}} ) = \pm \frac{L_0}{2\pi} \delta_{n,n'}\delta^2({\bm p}_{\perp} - {\bm p}'_{\perp})  \\
	\\
	\displaystyle ( {}_{\mp}\phi_{{\bm p}_{\perp},n}^{{\rm as}} | {}_{\pm}\phi_{{\bm p}'_{\perp},n'}^{{\rm as}} ) = 0
\end{array}\right.   \label{eq21}
\end{align}
with $(\phi_1|\phi_2)$ being the Klein-Gordon inner product defined as 
\begin{align}
	( \phi_1| \phi_2 ) \equiv {\rm i} \int {\rm d}^3{\bm x}\; \phi_1^* \overset{\leftrightarrow}{\partial}_{\tau} \phi_2.  
\end{align}

Imposing the standard canonical commutation relations onto $\hat{\phi}^{\rm as}$, one can quantize $\hat{a}^{\rm as}_{{\bm p}_{\perp},n} , \hat{b}^{\rm as}_{{\bm p}_{\perp},n} $ to obtain annihilation operators at the corresponding asymptotic times.  From the normalization condition (\ref{eq21}), the commutation relation for the annihilation operators read
\begin{align}
\left\{ \begin{array}{l}
	\displaystyle \frac{L_0}{2\pi}\delta_{n,n'} \delta^2({\bm p}_{\perp} - {\bm p}'_{\perp}) 
		= [a^{\rm as}_{{\bm p}_{\perp},n}, a^{{\rm as}\dagger}_{{\bm p}'_{\perp},n'}]
		= [b^{\rm as}_{{\bm p}_{\perp},n}, b^{{\rm as}\dagger}_{{\bm p}'_{\perp},n'}]  \\ \\
	\displaystyle ({\rm others}) = 0
\end{array}\right. ,
\end{align}
which, as usual, can be interpreted that $\hat{a}^{\rm as}_{{\bm p}_{\perp},n}$ and $\hat{b}^{\rm as}_{{\bm p}_{\perp},n}$ destroy a particle and an anti-particle with quantum number $n, {\bm p}_{\perp}$ at the corresponding asymptotic times, respectively.

The annihilation operators at in- and out-states are not independent with each other but are related with each other by a Bogoliubov transformation.  To see this, we use Eq.~(\ref{eq20}) and
\begin{align}
	&G_{\rm R}(x,x')
	= {\rm i}\theta(\tau - \tau') [ \phi^{\rm as}(x) , \phi^{{\rm as}\dagger}(x') ] \nonumber\\
	&= {\rm i}\theta(\tau - \tau') \frac{2\pi}{L_0} \sum_{n=1}^{\infty} \int {\rm d}^2{\bm p}_{\perp} \nonumber\\
	&\quad \times \left[    {}_{+}\phi_{{\bm p}_{\perp},n}^{{\rm as}}(x) {}_{+}\phi_{{\bm p}_{\perp},n}^{{\rm as}*}(x')  -   {}_{-}\phi_{{\bm p}_{\perp},n}^{{\rm as}}(x) {}_{-}\phi_{{\bm p}_{\perp},n}^{{\rm as}*}(x')  \right].  
\end{align}
Then, we find
\begin{align}
	&\begin{pmatrix} a^{\rm in/out}_{{\bm p}_{\perp},n} \\ b^{{\rm in/out}\dagger}_{-{\bm p}_{\perp},n} \end{pmatrix} 
	= \begin{pmatrix} ( {}_{+}\phi_{{\bm p}_{\perp},n}^{{\rm in/out}} | \phi^{\rm in/out} ) \\ - ( {}_{-}\phi_{{\bm p}_{\perp},n}^{{\rm in/out}} | \phi^{\rm in/out} ) \end{pmatrix} \nonumber\\
	&= \lim_{\tau \to -\infty/+\infty} \frac{1}{\sqrt{Z}} \begin{pmatrix} ( {}_{+}\phi_{{\bm p}_{\perp},n}^{{\rm in/out}} | \phi ) \\ - ( {}_{-}\phi_{{\bm p}_{\perp},n}^{{\rm in/out}} | \phi ) \end{pmatrix} \nonumber\\
	&=  \left( \begin{array}{l} 
			\displaystyle ({}_{+}\phi_{{\bm p}_{\perp},n}^{{\rm out}} | \phi^{\rm in} ) \\
			\displaystyle \quad + {\rm i} \int^{+\infty}_{-\infty} {\rm d}^4 y {}_+\phi^{{\rm out}*}_{{\bm p}_{\perp},n}(y) V(y) \phi^{\rm in} (y) + {\mathcal O} \left( \left(l/L_0 \right)^2 \right) \\
			\\
			\displaystyle - ( {}_{-}\phi_{{\bm p}_{\perp},n}^{{\rm out}} | \phi^{\rm in} ) \\
			\displaystyle \quad + {\rm i} \int^{+\infty}_{-\infty} {\rm d}^4 y {}_-\phi^{{\rm out}*}_{{\bm p}_{\perp},n}(y) V(y) \phi^{\rm in} (y)   + {\mathcal O} \left( \left(l/L_0 \right)^2 \right)
		\end{array} \right).  
\end{align}
Therefore, 
\begin{widetext}
\begin{subequations}
\begin{align}
	a^{\rm out}_{{\bm p}_{\perp},n}
		&= \frac{2\pi}{L_0}\sum_{n'}\int {\rm d}^2{\bm p}'_{\perp}  \left[ \left\{  ({}_{+}\phi_{{\bm p}_{\perp},n}^{{\rm out}} | {}_+\phi_{{\bm p}'_{\perp},n'}^{\rm in} ) + {\rm i} \int {\rm d}^4 y {}_+\phi^{{\rm out}*}_{{\bm p}_{\perp},n} V {}_+\phi_{{\bm p}'_{\perp},n'}^{\rm in} + {\mathcal O} \left( \left(l/L_0 \right)^2 \right)   \right\} a^{\rm in}_{{\bm p}'_{\perp},n'}  \right.\nonumber\\
			&\quad\quad\quad\quad\quad\quad\quad\quad\quad \left.  +   \left\{  ({}_{+}\phi_{{\bm p}_{\perp},n}^{{\rm out}} | {}_-\phi_{{\bm p}'_{\perp},n'}^{\rm in} ) + {\rm i} \int {\rm d}^4 y {}_+\phi^{{\rm out}*}_{{\bm p}_{\perp},n} V{}_-\phi_{{\bm p}'_{\perp},n'}^{\rm in} + {\mathcal O} \left( \left(l/L_0 \right)^2 \right)  \right\} b^{{\rm in}\dagger}_{-{\bm p}'_{\perp},n'}    \right], \\
	b^{{\rm out}\dagger}_{-{\bm p}_{\perp},n} 
		&= \frac{2\pi}{L_0}\sum_{n'}\int {\rm d}^2{\bm p}'_{\perp} \left[ \left\{ - ({}_{-}\phi_{{\bm p}_{\perp},n}^{{\rm out}} | {}_+\phi_{{\bm p}'_{\perp},n'}^{\rm in} ) + {\rm i} \int {\rm d}^4 y {}_-\phi^{{\rm out}*}_{{\bm p}_{\perp},n} V {}_+\phi_{{\bm p}'_{\perp},n'}^{\rm in} + {\mathcal O} \left( \left(l/L_0 \right)^2 \right)   \right\} a^{\rm in}_{{\bm p}'_{\perp},n'}   \right.\nonumber\\
			&\quad\quad\quad\quad\quad\quad\quad\quad\quad \left.  +     \left\{  - ({}_{-}\phi_{{\bm p}_{\perp},n}^{{\rm out}} | {}_-\phi_{{\bm p}'_{\perp},n'}^{\rm in} ) + {\rm i} \int {\rm d}^4 y {}_- \phi^{{\rm out}*}_{{\bm p}_{\perp},n} V{}_-\phi_{{\bm p}'_{\perp},n'}^{\rm in} + {\mathcal O} \left( \left(l/L_0 \right)^2 \right)  \right\} b^{{\rm in}\dagger}_{-{\bm p}'_{\perp},n'}    \right] .  
\end{align}
\end{subequations}
\end{widetext}
An important point here is that the annihilation operators at out-state $\hat{a}^{\rm out}_{{\bm p}_{\perp},n} , \hat{b}^{\rm out}_{{\bm p}_{\perp},n}$ contain the creation operators at in-state $\hat{a}^{{\rm in}\dagger}_{{\bm p}_{\perp},n} , \hat{b}^{{\rm in}\dagger}_{{\bm p}_{\perp},n}$ if the inner products and the matrix elements are non-vanishing.  It is easy to show that the matrix elements can be non-vanishing only when ${}_{\pm} \phi^{\rm in}_{{\bm p}_{\perp},n} \neq {}_{\pm} \phi^{\rm out}_{{\bm p}_{\perp},n}$.  This implies that the annihilation operators at out-state $\hat{a}^{\rm out}_{{\bm p}_{\perp},n} , \hat{b}^{\rm out}_{{\bm p}_{\perp},n}$ cannot annihilate the in-vacuum state if ${}_{\pm} \phi^{\rm in}_{{\bm p}_{\perp},n} \neq {}_{\pm} \phi^{\rm out}_{{\bm p}_{\perp},n}$ and, in turn, implies that the in-vacuum expectation value of the number operator at out-state becomes non-vanishing.

Now, we are ready to compute the total production number of particles $N$ and anti-particles $\bar{N}$ by directly evaluating the in-vacuum expectation value of the number operator at out-state $t=\tau \to +\infty$.  The result is
\begin{align}
	\overset{(-)}{N} 
		&\equiv S_\perp L_0 \times \frac{2\pi}{L_0}\sum_{n=1}^{\infty} \int {\rm d}^2{\bm p}_{\perp} \; \overset{(-)}{n}_{{\bm p}_{\perp},n} , \label{eq26}
\end{align}
where $S_\perp L_0$ is the volume of the system, and 
\begin{subequations}
\begin{align}
	n_{{\bm p}_{\perp},n}
		&\equiv \frac{1}{S_\perp L_0} \frac{   \braket{{\rm vac;in} | a^{{\rm out}\dagger}_{{\bm p}_{\perp},n}a^{{\rm out}}_{{\bm p}_{\perp},n}  | {\rm vac;in} }    }{ \braket{{\rm vac;in}  |  {\rm vac;in} } } \nonumber\\
		&=  \frac{1}{S_\perp L_0} \frac{2\pi}{L_0} \sum_{n'=1}^{\infty} \int {\rm d}^2{\bm p}'_{\perp} 
				\biggl| ({}_{+}\phi_{{\bm p}_{\perp},n}^{{\rm out}} | {}_-\phi_{{\bm p}'_{\perp},n'}^{\rm in} ) \nonumber\\
			&\quad + {\rm i} \int {\rm d}^4 y {}_+\phi^{{\rm out}*}_{{\bm p}_{\perp},n} V{}_-\phi_{{\bm p}'_{\perp},n'}^{\rm in} + {\mathcal O} \left( \left(l/L_0 \right)^2 \right)   \biggl|^2 , \label{eq36a} \\
	\bar{n}_{{\bm p}_{\perp},n} 
		&\equiv \frac{1}{S_\perp L_0}  \frac{   \braket{{\rm vac;in} | b^{{\rm out}\dagger}_{{\bm p}_{\perp},n} b^{{\rm out}}_{{\bm p}_{\perp},n}  | {\rm vac;in} }    }{ \braket{{\rm vac;in}  |  {\rm vac;in} } } \nonumber\\
		&=  \frac{1}{S_\perp L_0} \frac{2\pi}{L_0} \sum_{n'=1}^{\infty} \int {\rm d}^2{\bm p}'_{\perp} 
				\biggl| - ({}_{-}\phi_{{\bm p}_{\perp},n}^{{\rm out}} | {}_+ \phi_{{\bm p}'_{\perp},n'}^{\rm in} ) \nonumber\\
			&\quad + {\rm i} \int {\rm d}^4 y {}_-\phi^{{\rm out}*}_{{\bm p}_{\perp},n} V {}_+\phi_{{\bm p}'_{\perp},n'}^{\rm in} + {\mathcal O} \left( \left(l/L_0 \right)^2 \right) \biggl|^2 ,  \label{eq36b} 
\end{align}
\end{subequations}
are the distributions of particles and anti-particles for each mode ${\bm p}_{\perp},n$, respectively.  The in-vacuum state at $t=\tau \to -\infty$, which we write $\ket{{\rm vac};{\rm in}}$, is defined as
\begin{align}
	0 = a^{{\rm in}}_{{\bm p}_{\perp},n} \ket{{\rm vac};{\rm in}} = b^{{\rm in}}_{{\bm p}_{\perp},n} \ket{{\rm vac};{\rm in}}\ {\rm for\ any}\ {\bm p}_{\perp},n.   
\end{align}
Note that $n',{\bm p}'_{\perp}$ in Eqs.~(\ref{eq36a}) and (\ref{eq36b}) can be interpreted as the quantum number of the pairly produced particle.  The matrix element in Eqs.~(\ref{eq36a}) and (\ref{eq36b}) always has off-diagonal components $n \neq n'$ because the vibration gives finite longitudinal momentum to the produced particles.

The production number formula (\ref{eq26}) smoothly connects the non-perturbative Schwinger mechanism and the parametric dynamical Casimir effect: In the limit of no vibration $V \to 0$, the second term in Eqs.~(\ref{eq36a}) and (\ref{eq36b}) vanishes.  Then, we only have the first term, which describes the particle production by an electric field.  This is, by definition, the Schwinger mechanism.  On the other hand, the second term includes the effect of the vibration.  In the absence of an electric field ${\bm E}_{\perp} \to 0$ (or, equivalently, $V$ becomes much greater than ${\bm E}_{\perp}$), our formalism reduces to the standard perturbation theory without ${\bm E}_{\perp}$, in which the first term vanishes and only the second term survives.  Then, the particle production is driven by the vibration, which is nothing but the dynamical Casimir effect.  For general ${\bm E}_{\perp}$ and/or $V$, our formula (\ref{eq26}) deviates from the standard formula for the Schwinger mechanism or the dynamical Casimir effect because (i) we have an interference between the first and second terms; and (ii) our mode function ${}_{\pm}\phi_{{\bm p}_{\perp},n}^{\rm as}$ is fully dressed by the electric field ${\bm E}_{\perp}$ and hence the second term is strongly affected by $e$ and ${\bm E}_{\perp}$ in a non-linear manner.  Note that the above argument suggests that the typical frequency of the vibration can control the interplay between the Schwinger mechanism and the dynamical Casimir effect.  This is because the strength of the potential $V$ (\ref{eq16}) increases with the frequency, with which one can control the relative strength between ${\bm E}_{\perp}$ and $V$.  We explicitly confirm this expectation in the next section.

\section{Constant and homogeneous electric field} \label{sec3}

We discuss quantitative aspects of the particle production based on the general formula obtained in Sec.~\ref{sec2}.  As a demonstration, let us assume that the electric field ${\bm E}_{\perp}$ is sufficiently slow such that it is well approximated by a constant and homogeneous electric field, i.e., 
\begin{align}
	{\bm E}_{\perp} (x) = E {\bf e}_x\ {\rm with}\ eE>0.  \label{eq-28}
\end{align}
An advantage of this field configuration is that the field equation (\ref{eq25}) is analytically solvable, with which one can exactly evaluate the production number (\ref{eq26}).  Also, the Schwinger mechanism for the constant and homogeneous electric field configuration is very well understood \cite{sch51}, while that for a spacetime-dependent electric field configuration is less understood.  It is, therefore, good to consider the well-understood field configuration, so that we can better understand physics of the dynamical assistance by the dynamical Casimir effect to the Schwinger mechanism and vice versa on a clear footing.  

The rest of this section is organized as follows: In Sec~\ref{sec3A}, we present the analytical expression for the production number (\ref{eq26}) using the analytical solution of the field equation (\ref{eq25}).  In Sec.~\ref{sec3b}, we analytically discuss how the particle production in the presence of both a strong electric field and vibrating plates is related to the standard Schwinger mechanism and the dynamical Casimir effect based on the analytical expression for the production number obtained in Sec.~\ref{sec3A}.  We also show that the interplay between the two production mechanisms can be controlled by the typical frequency of the vibration.  In Sec.~\ref{sec3c}, we numerically discuss the dynamical assistance between the two production mechanisms and show that it dramatically enhances the particle production number compared to what the standard Schwinger mechanism or the dynamical Casimir effect naively expects.

\subsection{Evaluation of the formula (\ref{eq26})} \label{sec3A}

Let us evaluate the production number formula (\ref{eq26}) for the constant and homogeneous electric field configuration (\ref{eq-28}).  For this field configuration, one can analytically solve the mode equation (\ref{eq25}) as
\begin{align}
	\begin{pmatrix} {}_{+}\phi_{{\bm p}_{\perp},n}^{{\rm as}}(x) \\ {}_{-}\phi_{{\bm p}_{\perp},n}^{{\rm as}}(x) \end{pmatrix}
	= \begin{pmatrix} \Phi_{{\bm p}_{\perp},n}^{{\rm as}} (\tau) \\ \Phi_{{\bm p}_{\perp},n}^{{\rm as}*}(\tau) \end{pmatrix} 
		\frac{{\rm e}^{i{\bm p}_{\perp}\cdot{\bm x}_{\perp}}}{2\pi} \sqrt{\frac{L_0}{\pi}} \sin (n\pi \xi),\label{eq29}
\end{align}
where
\begin{align}
\left\{ \begin{array}{l}
	\displaystyle \Phi_{{\bm p}_{\perp},n}^{{\rm in}}  
	= \frac{{\rm e}^{-\pi a_{{\bm p}_{\perp},n}/4}}{(2eE)^{1/4}} \\
	\quad\quad\quad\quad \times \left[ D_{-{\rm i}a_{{\bm p}_{\perp},n}-1/2} \left( -{\rm e}^{{\rm i}\pi/4} \sqrt{\frac{2}{eE}}(eE\tau + p_x) \right)  \right]^* \\ \\
	\displaystyle \Phi_{{\bm p}_{\perp},n}^{{\rm out}} 
	= \frac{{\rm e}^{-\pi a_{{\bm p}_{\perp},n}/4}}{(2eE)^{1/4}} \\
	\quad\quad\quad\quad \times D_{-{\rm i}a_{{\bm p}_{\perp},n}-1/2}\left( {\rm e}^{{\rm i}\pi/4} \sqrt{\frac{2}{eE}}(eE\tau + p_x) \right)  
\end{array} \right. . \label{eqa2}
\end{align}
Here, $D_{\nu}(z)$ is the parabolic cylinder function and 
\begin{align}
	a_{{\bm p}_{\perp},n} \equiv \frac{m^2 + p_y^2 + \left( \frac{n \pi}{L_0} \right)^2}{2eE} .  
\end{align}

Using the solution (\ref{eq29}), one can exactly evaluate the inner product and the matrix element in Eqs.~(\ref{eq36a}) and (\ref{eq36b}).  We find
\begin{align}
	&({}_{+}\phi_{{\bm p}_{\perp},n}^{{\rm out}} | {}_- \phi_{{\bm p}'_{\perp},n'}^{\rm in} ) = - [({}_{+}\phi_{{\bm p}_{\perp},n}^{{\rm out}} | {}_- \phi_{{\bm p}'_{\perp},n'}^{\rm in} )]^*\nonumber\\ 
	&= \delta^2({\bm p}_{\perp} - {\bm p}'_{\perp}) \times \frac{L_0}{2\pi}\delta_{n,n'} \times (+{\rm i}){\rm e}^{-\pi a_{{\bm p}_{\perp},n}}, \label{eq-32}
\end{align}
and
\begin{align}
	&{\rm i} \int {\rm d}^4 y {}_{\pm} \phi^{{\rm out}*}_{{\bm p}_{\perp},n} V{}_{\mp}\phi_{{\bm p}'_{\perp},n'}^{\rm in} \nonumber\\
	&= \delta^2({\bm p}_{\perp} - {\bm p}'_{\perp}) \times \frac{L_0}{2\pi} \times {\rm i} \frac{n\pi}{\sqrt{eE}L_0}\frac{n'\pi}{\sqrt{eE}L_0} (-1)^{n+n'}\nonumber\\
	&\quad \times \int^{\infty}_{0} {\rm d}\omega \frac{\tilde{l}(\omega)}{L_0} {\rm e}^{{\rm i} \frac{\omega^2 - 4p_x \omega}{4eE}}\left( \frac{\omega}{\sqrt{2eE}}    \right)^{ {\rm i}(a_{{\bm p}_{\perp},n} - a_{{\bm p}_{\perp},n'}) } \nonumber\\
	&\quad \times {}_1 \tilde{F}_1 \left( \frac{1}{2} + {\rm i} a_{{\bm p}_{\perp},n}  ; 1+{\rm i}(a_{{\bm p}_{\perp},n} - a_{{\bm p}_{\perp},n'}) ; -{\rm i}\frac{\omega^2}{2eE}   \right) , \label{eq-33}
\end{align}
where 
\begin{align}
	\tilde{l}(\omega) \equiv \int^{+\infty}_{-\infty}{\rm d}t\;{\rm e}^{-{\rm i}\omega t}l(t),
\end{align}
and ${\mathcal O}((l/L_0)^2)$-term is discarded.  Putting Eqs.~(\ref{eq-32}) and (\ref{eq-33}) into Eqs.~(\ref{eq36a}) and (\ref{eq36b}), we arrive at
\begin{align}
	&n_{{\bm p}_{\perp},n} = \bar{n}_{-{\bm p}_{\perp},n} \nonumber\\
	&= \frac{1}{(2\pi)^3} \sum_{n'=1}^{\infty} {\rm e}^{-\pi (a_{{\bm p}_{\perp},n} + a_{{\bm p}_{\perp},n'}) } \Biggl| 
				\delta_{n,n'}+ \frac{n \pi}{\sqrt{eE}L_0}\frac{n'\pi}{\sqrt{eE}L_0} \nonumber\\
				&\quad \times (-1)^{n+n'}\int^{\infty}_{0} {\rm d}\omega \frac{\tilde{l}(\omega)}{L_0} {\rm e}^{{\rm i} \frac{\omega^2 - 4p_x \omega}{4eE}}\left( \frac{\omega}{\sqrt{2eE}}    \right)^{ {\rm i}(a_{{\bm p}_{\perp},n} - a_{{\bm p}_{\perp},n'}) } \nonumber\\
				&\quad \times{}_1 \tilde{F}_1 \left( \frac{1}{2} + {\rm i} a_{{\bm p}_{\perp},n}  ; 1+{\rm i}(a_{{\bm p}_{\perp},n} - a_{{\bm p}_{\perp},n'}) ; -{\rm i}\frac{\omega^2}{2eE}   \right) 
		\Biggl|^2 . \label{eq--36}
\end{align}

The total production number (\ref{eq26}) can be evaluated by integrating Eq.~(\ref{eq--36}).  By noting $\int {\rm d}p_x = eE T$ with $T$ being the whole time interval \cite{nik70, tan09}, we find
\begin{widetext}
\begin{align}
	N=\bar{N} 
	&= \frac{S_{\perp}L_0T}{(2\pi)^3} m^4 \sum_n \sum_{n'}  2 \frac{\pi}{\sqrt{eE}L_0} \left| \frac{eE}{m^2}  \right|^2 {\rm e}^{-\pi \frac{m^2}{eE}} \Biggl[
			\delta_{n,n'} \times {\rm e}^{-\pi \left( \frac{n \pi}{\sqrt{eE}L_0} \right)^2} \left\{ 1 + \frac{2\pi}{T} \left| \frac{n \pi}{\sqrt{eE}L_0} \right|^2 \frac{\tilde{l}(0)}{L_0}  \right\} \nonumber\\
			&\quad
			+ \frac{2\pi}{T}\frac{1}{\sqrt{eE}} \left| \frac{n \pi}{\sqrt{eE}L_0} \right|^2 \left| \frac{n' \pi}{\sqrt{eE}L_0} \right|^2 {\rm e}^{- \pi \frac{n^2+{n'}^2}{2} \left| \frac{\pi}{\sqrt{eE}L_0} \right|^2} \int {\rm d}p_y \int_0^{\infty} {\rm d}\omega \;{\rm e}^{-\pi\frac{p_y^2}{eE}} \left| \frac{\tilde{l}(\omega)}{L_0}\right|^2\nonumber\\
				&\quad\quad \times 
				 \left| {}_1 \tilde{F}_1 \left( \frac{1}{2} + \frac{{\rm i}}{2} \left\{  \frac{m^2+p_y^2}{eE} +  \left| \frac{n\pi}{\sqrt{eE}L_0} \right|^2  \right\}; 1 + {\rm i} \frac{n^2-{n'}^2}{2}  \left| \frac{\pi}{\sqrt{eE}L_0} \right|^2 ; - {\rm i}\frac{\omega^2}{2eE}   \right)  \right|^2 \Biggl]. \label{eq-36}
\end{align}
\end{widetext}

\subsection{The interplay between the Schwinger mechanism and the dynamical Casimir effect} \label{sec3b}

Let us discuss the basic features of the production number formula (\ref{eq-36}).  Namely, we analytically discuss how the formula (\ref{eq-36}) is related to the Schwinger mechanism and the dynamical Casimir effect and how the interplay between the two production mechanisms can be described in terms of the typical frequency of the vibration.

\subsubsection{Slow vibration: the Schwinger mechanism}

For a slow vibration, the plates cannot supply large energy to the vacuum.  Therefore, the main energy source for the particle production should come from the electric field.  This implies that the particle production is dominated by the Schwinger mechanism, for which the production number is exponentially suppressed by the mass gap.

To see this, let $\Omega$ be the typical frequency of the vibration and assume $\Omega \sim 0$.  This is equivalent to assuming 
\begin{align}
	l \sim l_0\ \Leftrightarrow\ \tilde{l} \sim 2\pi \delta(\omega) \times l_0.  \label{eq--38} 
\end{align}
Then, one can simplify the distribution $\overset{(-)}{n}_{{\bm p}_{\perp},n}$ (\ref{eq--36}) as
\begin{align}
	&{n}_{{\bm p}_{\perp},n} = \bar{n}_{-{\bm p}_{\perp},n} \nonumber\\
	&\xrightarrow{\Omega \to 0}
			\frac{1}{(2\pi)^3} \left[ 1 + 2 \pi \left( \frac{n\pi}{\sqrt{eE}L_0} \right)^2 \frac{l_0}{L_0} + {\mathcal O}\left( (l_0/L_0)^2 \right)   \right] \nonumber\\
			&\quad\quad\quad \times \exp\left[ -\pi \frac{m^2 + p_y^2 + ( n\pi/L_0)^2 }{eE} \right] \nonumber\\
	&= \frac{1}{(2\pi)^3} \exp\left[ -\pi \frac{m^2 + p_y^2 + ( n\pi/L)^2 }{eE} \right] \left[ 1 + {\mathcal O}\left( (l_0/L_0)^2 \right)   \right], \label{eq__39}
\end{align}
where
\begin{align}
	L = L_0 + l_0 
\end{align}
is the total size of the system (\ref{eq--2}) in the slow limit (\ref{eq--38}).  Up to ${\mathcal O}((l_0/L_0)^2)$, Eq.~(\ref{eq__39}) reproduces the well-known Schwinger formula \cite{sch51} for a system with size $L$, in which the longitudinal momentum $p_z$ is quantized as $p_z \to n\pi/L$.  Therefore, the particle production is, indeed, dominated by the Schwinger mechanism for a slow vibration.  It should be noted that the finiteness of the system always makes the mass gap $\omega_{{\bm p}_{\perp},n}$ larger by $\sim n\pi/L$ and hence the particle production is always suppressed if the frequency is small, i.e., if the production is dominated by the Schwinger mechanism.

The total production number $\overset{(-)}{N}$ can be evaluated by integrating Eq.~(\ref{eq__39}).  By neglecting terms of the order of ${\mathcal O}((l_0/L_0)^2)$, we find
\begin{align}
	\overset{(-)}{N} 
	\xrightarrow{\Omega \to 0}&
	\frac{S_{\perp} L T}{(2\pi)^3} m^4 \left|\frac{eE}{m^2}\right|^2 {\rm e}^{-\pi \frac{m^2}{eE}} \nonumber\\
	&\times \frac{\pi}{\sqrt{eE}L} \left[  -1 + \vartheta_3\left( 0, {\rm e}^{-\pi \left( \frac{\pi}{\sqrt{eE}L} \right)^2} \right)  \right] \nonumber\\
	\equiv& \overset{(-)}{N}_{({\rm Sch}; L \to \infty)} \times F \left( \frac{\pi}{\sqrt{eE}L}  \right) \nonumber\\
	\equiv& \overset{(-)}{N}_{({\rm Sch})}, \label{eq39}
\end{align}
where $\vartheta_3$ is the Jacobi elliptic function of the third kind, and $\overset{(-)}{N}_{({\rm Sch};L\to\infty)} $ is the Schwinger formula for the total production number with infinite system size $L \to \infty$ \cite{sch51}, 
\begin{align}
	\overset{(-)}{N}_{({\rm Sch}; L\to \infty)} 
	&\equiv S_{\perp} L \int \frac{{\rm d}^3{\bm p}}{(2\pi)^3} \exp \left[ -\pi \frac{m^2 + p_x^2 + p_z^2}{eE} \right] \nonumber\\
	&= \frac{S_{\perp} L T}{(2\pi)^3} m^4 \left|\frac{eE}{m^2}\right|^2 {\rm e}^{-\pi \frac{m^2}{eE}}. \label{eq---42}
\end{align}
The factor $F \leq 1$ in Eq.~(\ref{eq39}) accounts for the finite size effect, which suppresses the Schwinger mechanism as 
\begin{align}
	F \left( \frac{\pi}{\sqrt{eE}L} \right) 
	\to \left\{ 
			\begin{array}{ll} 
				1 - \frac{\pi}{\sqrt{eE}L}  &  {\rm for}\ \frac{\pi}{\sqrt{eE}L}\ll 1 \\ 
				2 \frac{\pi}{\sqrt{eE}L} {\rm e}^{-\pi \left( \frac{\pi}{\sqrt{eE}L} \right)^{2}}  &  {\rm for}\ \frac{\pi}{\sqrt{eE}L}\gg 1 
			\end{array} \right. .
\end{align}
Namely, the finite size effect is not important for large $L \gg 1/\sqrt{eE}$, while it gives an exponential suppression for small $L \ll 1/\sqrt{eE}$ for which the mass gap $\omega_{\bm p} = \sqrt{m^2+p_y^2+(n\pi/L)^2}$ becomes larger than the electric field strength no matter how small $m$ is.  Note that the finite transverse size effect to the Schwinger mechanism was previously discussed in Refs.~\cite{ceh95, pav91, sch90, cec88, cec89, gat92, war92} (with various theoretical approaches and boundary conditions), whose results are consistent with ours.

\subsubsection{Fast vibration: the dynamical Casimir effect}

If the vibration is fast enough, the plates are able to supply large energy to the vacuum.  Then, the particle production should be driven by the vibrating plates rather than by the electric field.  Therefore, the dynamical Casimir effect should dominate the production, for which the production number is just suppressed by powers of the mass gap and is free from the strong exponential suppression unlike the Schwinger mechanism.

To see this, let us assume that the vibration is dominated by a high-frequency mode $\Omega \to \infty$, for which the distribution $\overset{(-)}{n}_{{\bm p}_{\perp},n}$ (\ref{eq--36}) reads
\begin{align}
	&{n}_{{\bm p}_{\perp},n} = \bar{n}_{-{\bm p}_{\perp},n} \nonumber\\
	&\xrightarrow{\Omega \to \infty}
			\frac{1}{(2\pi)^3} \sum_{n'} \Biggl| \delta_{n,n'} \exp\Biggl[ -\pi \frac{m^2 + p_y^2 + ( n\pi/L_0 )^2 }{eE} \Biggl] \nonumber\\
	&\quad\quad\quad + \frac{(-1)^{n+n'}}{\sqrt{\omega_{{\bm p}_{\perp},n}\omega_{{\bm p}_{\perp},n'}}} \frac{n\pi}{L_0}\frac{n'\pi}{L_0} \frac{\tilde{l}^*(\omega_{{\bm p}_{\perp},n}+\omega_{{\bm p}_{\perp},n'})}{L_0} \Biggl|^2 \nonumber\\
	&\ \quad \sim \frac{1}{(2\pi)^3} \sum_{n'} \frac{1}{\omega_{{\bm p}_{\perp},n}\omega_{{\bm p}_{\perp},n'}}\nonumber\\
	&\ \quad\quad\times \left( \frac{n\pi}{L_0} \right)^2 \left( \frac{n'\pi}{L_0} \right)^2  \left| \frac{\tilde{l}(\omega_{{\bm p}_{\perp},n}+\omega_{{\bm p}_{\perp},n'})}{L_0} \right|^2. \label{eq--43}
\end{align}
Therefore, we obtain
\begin{align}
	\lim_{\Omega \to \infty} \overset{(-)}{N}
	&\sim \frac{S_\perp T}{(2\pi)^3} \sum_n\sum_{n'} \int {\rm d}p_y \frac{1}{\omega_{{\bm p}_{\perp},n}\omega_{{\bm p}_{\perp},n'}} \nonumber\\
	&\quad\times \left( \frac{n\pi}{L_0} \right)^2 \left( \frac{n'\pi}{L_0} \right)^2\left| \frac{\tilde{l}(\omega_{{\bm p}_{\perp},n}+\omega_{{\bm p}_{\perp},n'})}{L_0} \right|^2 \nonumber \\
	&\equiv \overset{(-)}{N}_{({\rm Cas})}.   \label{eq--44}
\end{align}
Equations~(\ref{eq--43}) and (\ref{eq--44}) are independent of the electric field $E$ and are dependent only on the vibration $l$, and reproduce the known formulas for the dynamical Casimir effect \cite{mun98}.  Therefore, the particle production is, indeed, dominated by the dynamical Casimir effect for a fast vibration.  Note that, to get the second line of Eq.~(\ref{eq--43}), we neglected the first term, which comes from the inner product in Eqs.~(\ref{eq36a}) and (\ref{eq36b}), because it is exponentially suppressed by the mass gap $\omega_{{\bm p}_{\perp},n}$.  In other words, the contribution from the first term (i.e., the Schwinger mechanism) is subleading compared to the second term (i.e., the dynamical Casimir effect), which is just suppressed by powers of the mass gap $\omega_{{\bm p}_{\perp},n}$.  For a very strong electric field comparable to the mass gap $eE \gtrsim \omega_{{\bm p}_{\perp},n}^2$, the first term (i.e., the Schwinger mechanism) is free from the exponential suppression, for which case  one may not neglect it.  Note also that the argument of $\tilde{l}$ is $\omega_{{\bm p}_{\perp},n}+\omega_{{\bm p}_{\perp},n'}$, which appears because the vibration must supply energy $\omega_{{\bm p}_{\perp},n}+\omega_{{\bm p}_{\perp},n'}$ to create a pair of particles with energy $\omega_{{\bm p}_{\perp},n}$ and $\omega_{{\bm p}_{\perp},n'}$.  This implies that the (lowest-order parametric) dynamical Casimir effect never occurs for vibrations whose typical frequency $\Omega$ is below $\omega_{{\bm p}_{\perp},n}+\omega_{{\bm p}_{\perp},n'}$\footnote{Note that our discussion here is based on the first order perturbation theory with respect to the vibration, which means that the vibration can interact with particles only once.  If one considers higher $n$-th order perturbations, i.e., multiple interactions between the vibration and particles, the threshold frequency can be lowered as $\Omega = (\omega_{{\bm p}_{\perp},n}+\omega_{{\bm p}_{\perp},n'})/n$.  Such higher order perturbations are, however, suppressed by $(l_0/L_0)^n$ and hence can safely be neglected.  }.

\subsection{The dynamical assistance} \label{sec3c}

At intermediate frequencies (i.e., the vibration is neither fast or slow), both the Schwinger mechanism and the dynamical Casimir effect become important, and they assist with each other to enhance the particle production.  

We demonstrate how the assistance occurs by explicitly evaluating the total production number $N$ (\ref{eq-36}) by considering a monochromatic vibration as an example: 
\begin{align}
	l = l_0 \sin \Omega t, \label{eq_-45}
\end{align}
for which the Fourier component $\tilde{l}$ reads
\begin{align}
	\tilde{l} = l_0 \times {\rm i}\pi \left[ \delta(\omega+\Omega) -  \delta(\omega-\Omega) \right] .  \label{eq--46}
\end{align}

\subsubsection{$\Omega$-dependence}

\begin{figure}[!t]
\begin{center}
\includegraphics[clip, width=0.49\textwidth]{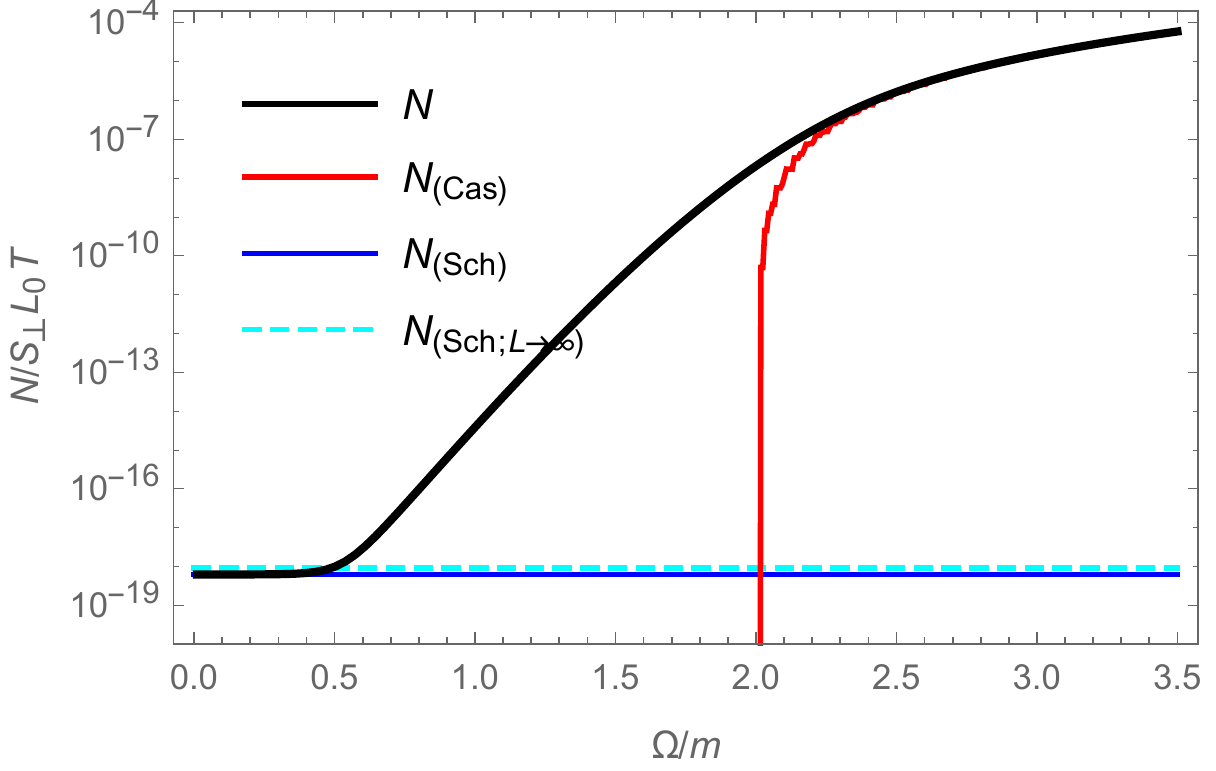}
\caption{\label{fig2} $\Omega$-dependence of the total production number $N$ (\ref{eq-36}) for the monochromatic vibration (\ref{eq_-45}).  As a comparison, the production number for the Schwinger mechanism with finite $L$ (\ref{eq39}) and with infinite $L\to \infty$ (\ref{eq---42}), and for the dynamical Casimir effect (\ref{eq--44}) are shown in blue, dashed cyan, and red lines, respectively.  The parameters are fixed as $eE/m^2=0.1, mL_0=30$, and $ml_0=0.3$.  }
\end{center}
\end{figure}

\begin{figure}[!t]
\begin{center}
\includegraphics[clip, width=0.49\textwidth]{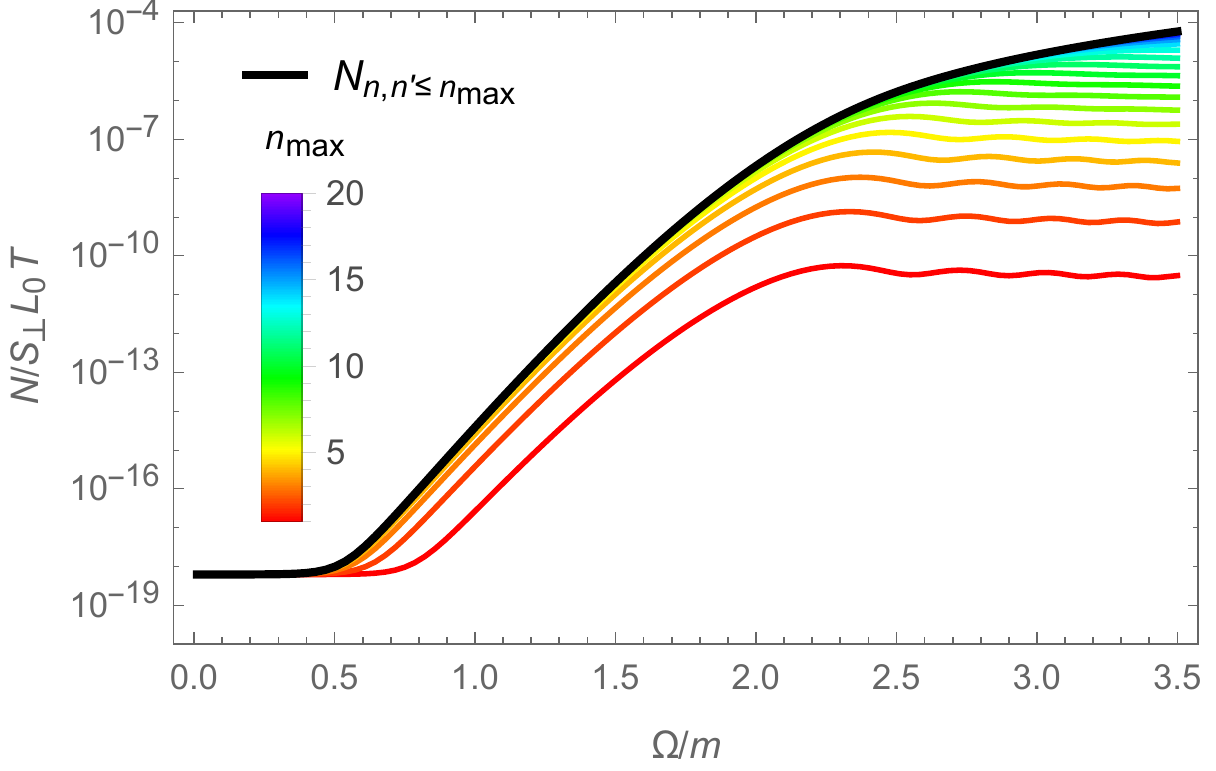}
\caption{\label{fig3} The contribution from the $n,n'$-th modes to the total production number $N$ (\ref{eq-36}) for the monochromatic vibration (\ref{eq_-45}).  Different colors distinguish the maximum value of the summation $n,n' \leq n_{\rm max} = 0,1,\ldots, 20$, and the black curve is for $n_{\rm max}=25$.  The parameters are the same as Fig.~\ref{fig2}, i.e., $eE/m^2=0.1, mL_0=30$, and $ml_0=0.3$.  }
\end{center}
\end{figure}

Figure~\ref{fig2} shows the result for the total production number $N$ (\ref{eq-36}) as a function of the frequency $\Omega$.  As a demonstration, we considered a subcritical electric field $eE/m^2=0.1$, sufficiently large system size $mL_0=30$, and a small vibration $ml_0=0.3$ (i.e., $l_0/L_0=1/100$).

Figure~\ref{fig2} clearly shows the interplay between the Schwinger mechanism and the dynamical Casimir effect that we discussed in Sec.~\ref{sec3b}.  Namely, the Schwinger mechanism (the dynamical Casimir effect) dominates the particle production when the frequency is small $\Omega \ll m,\sqrt{eE}, 1/L_0$ (large $\Omega \gg m,\sqrt{eE}, 1/L_0$), for which the production number is strongly (weakly) suppressed by an exponential (powers) of the mass gap $\omega_{{\bm p}_{\perp},n}$.  Notice that the dynamical Casimir effect occurs only above $|\Omega| \geq 2\sqrt{m^2 + (\pi/L_0)^2}>2m$ because of the energy conservation as we discussed below Eq.~(\ref{eq--44}).  This point may become clearer if one carry out the $p_y$-integration in Eq.~(\ref{eq--44}), which is analytically doable for the monochromatic vibration (\ref{eq_-45}) as
\begin{align}
	\overset{(-)}{N}_{({\rm Cas})} 
	&= \frac{S_\perp L_0 T}{(2\pi)^3} m^4 \times 2\pi^3 \left|\frac{eE}{m^2} \right|^2\left| \frac{l_0}{L_0} \right|^2 \nonumber\\
	&\quad \times \sum_n \sum_{n'} \left| \frac{n\pi}{L_0} \right|^2 \left| \frac{n'\pi}{L_0} \right|^2 \frac{1}{L_0 |\Omega|} \nonumber\\
	&\quad\times \Theta\left(| \Omega| - \sqrt{m^2+\left| \frac{n\pi}{L_0} \right|^2}- \sqrt{m^2+\left| \frac{n'\pi}{L_0} \right|^2} \right).   \label{e-q48}
\end{align}
Hence, if the production is dominated by the dynamical Casimir effect, the $n,n'$-th modes can contribute to the production only when $|\Omega| \geq \sqrt{m^2+(n\pi/L_0)^2} + \sqrt{m^2+(n'\pi/L_0)^2}$ is satisfied.  This is in contrast to the production by the Schwinger mechanism, which does not have such a sharp threshold behavior for the modes $n,n'$ because an electric field can mix up different energy states.

At intermediate frequencies (cf. $\Omega/m \sim 0.5$ in Fig.~\ref{fig2}), the production number $N$ is dramatically enhanced by orders of the magnitude compared to the naive expectations of the Schwinger mechanism and the dynamical Casimir effect.  This is the dynamical assistance effect between the two production mechanisms.  Note that the size of the enhancement changes quadratically with $l$ and also is dependent on the system size $L_0$ and the electric field strength $E$, which will be discussed in Sec.~\ref{sec3b2} and \ref{sec3b3}, respectively.  One may interpret this assistance in two different ways (although the physics is the same) depending on which mechanism one compares with the result: (i) Since the production number in the presence of the vibrating plates is much more abundant than the naive Schwinger formula without vibrating plates, one may say that the Schwinger mechanism is enhanced by the dynamical Casimir effect.  This may be understood as an analog of the dynamically assisted Schwinger mechanism not by a superposition of an additional electromagnetic field \cite{sch08, piz09, dun09, mon10a, mon10b} but by the vibrating plates.  Physically, the energy supply from the vibrating plates reduces the mass gap, and hence it becomes easier for the quantum tunneling by a strong electric field (i.e., the Schwinger mechanism) to occur.  (ii) The particle production occurs even below the threshold frequency for the lowest mode $|\Omega| \geq \sqrt{m^2+(n\pi/L_0)^2} + \sqrt{m^2+(n'\pi/L_0)^2}$.  Since the naive (lowest-order parametric) dynamical Casimir effect alone cannot create particles if the frequency is below the threshold, one may say that the Schwinger mechanism assists the dynamical Casimir effect to lower the threshold frequency.  Physically, the energy supplied by the electric field reduces the mass gap, which assists the perturbative pair production by the vibrating plates (i.e., the dynamical Casimir effect).  This may be understood as an analog of the Franz-Keldysh effect \cite{fra58, kel58, tah63, cal63, tay19}, in which photo-absorption rate in the presence of a strong electric field becomes finite even below the gap energy.

Figure~\ref{fig3} shows how the $n,n'$-th modes contribute to the production number $N$.       
The lowest mode $n,n'=1$ has the smallest mass gap, so that it gives the largest contribution to the production number.  In particular, the lowest mode dominates the production if the system size is very small $L_0 \ll |\Omega|^{-1}, |eE|^{-1/2}, m^{-1}$ (cf. the lowest Landau approximation in a strong magnetic field), for which the mass gap of the higher modes becomes very large and their production is strongly suppressed.  With increasing $\Omega$ (i.e., injecting more energy to the vacuum), higher $n,n'$-th modes begin to contribute at around the naive threshold frequency $|\Omega| \sim \sqrt{m^2+(n\pi/L_0)^2} + \sqrt{m^2+(n'\pi/L_0)^2}$.  Notice that there is no sharp threshold behavior as the dynamical Casimir effect expects (\ref{e-q48}) and the particle production has a tail below the threshold.  This is nothing but the dynamical assistance by the Schwinger mechanism to the dynamical Casimir effect that we discussed in the last paragraph.  Also, note that the production number for each mode exhibits an oscillating behavior as a function of $\Omega$ above the threshold.  This can be understood as an analog of the Franz-Keldysh oscillation \cite{tah63, cal63, tay19}, which occurs because of the quantum reflection process (a dual process of the quantum tunneling) in the presence of a strong electric field.  If $L_0$ is large such that higher $n,n'$-modes can contribute to the total production number $N$, it is hard to see the oscillation behavior in $N$ because the oscillation at each mode cancels with each other after the $n,n'$-summation.  Inversely, the oscillating behavior in $N$ may survive if $L_0$ is small such that only a few modes can contribute to the production number $N$.

\subsubsection{$L_0$-dependence}\label{sec3b2}

\begin{figure}[!t]
\begin{center}
\includegraphics[clip, width=0.49\textwidth]{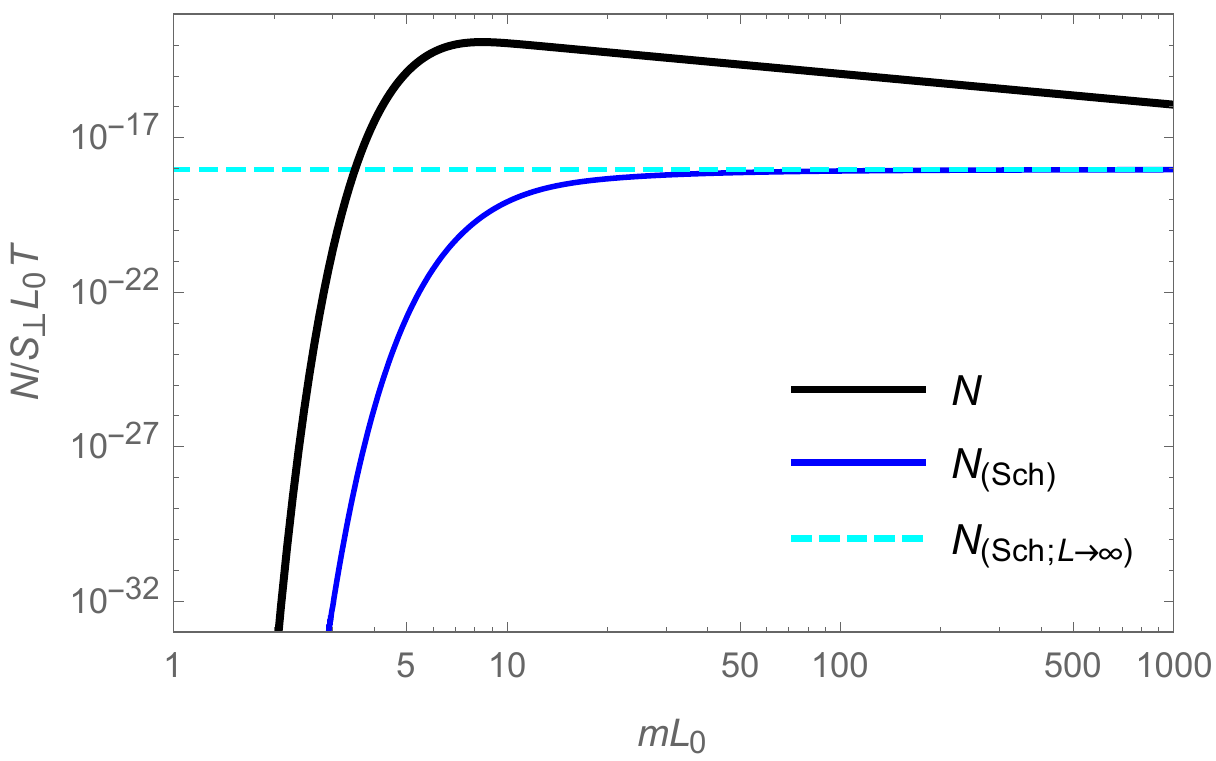}
\caption{\label{fig4} $L_0$-dependence of the total production number $N$ (\ref{eq-36}) for the monochromatic vibration (\ref{eq_-45}).  As a comparison, the production number for the Schwinger mechanism with finite $L$ (\ref{eq39}) and with infinite $L\to \infty$ (\ref{eq---42}) are shown in blue and dashed cyan, respectively.  The parameters are fixed as $eE/m^2=0.1, \Omega/m=1$, and $ml_0=0.3$.  }
\end{center}
\end{figure}

Figure~\ref{fig4} shows the system size $L_0$-dependence of the total production number $N$ (\ref{eq-36}).  As a demonstration, we considered a subcritical electric field $eE/m^2=0.1$ and an intermediate frequency $\Omega/m=1$.  We fixed the amplitude of the vibration as $ml_0=0.3$, and therefore the ratio $l_0/L_0$ decreases as increasing $L_0$.  Note that the naive dynamical Casimir effect expects no particle production $N_{\rm (Cas)}=0$ [Eq.~(\ref{eq--44})] for $\Omega/m=1$.

For small $L_0$, the particle production is strongly suppressed because of the finite size effect.  Although the dynamical assistance enhances the production by orders of the magnitude compared to the naive Schwinger formula with finite $L_0$, the finite size effect is so large that the production number becomes much smaller than the naive Schwinger formula with infinite $L_0 \to \infty$.  Indeed, one may analytically take the $L_0 \to 0$ limit of Eq.~(\ref{eq-36}) to find
\begin{align}
	&\lim_{L_0 \to 0} \overset{(-)}{N} \nonumber\\
	&= \frac{S_{\perp}L_0T}{(2\pi)^3} m^4 \times 2 \frac{\pi}{\sqrt{eE}L_0} \left| \frac{eE}{m^2}  \right|^2 {\rm e}^{-\pi \frac{m^2}{eE}} {\rm e}^{-\pi \left( \frac{\pi}{\sqrt{eE}L_0} \right)^2} \nonumber\\
	&\quad \times \Biggl[
			1 + \frac{2\pi}{T} \left| \frac{\pi}{\sqrt{eE}L_0} \right|^2 \frac{\tilde{l}(0)}{L_0} \nonumber\\
			&\quad\quad\ 
			+ \frac{2\pi}{T} \left| \frac{ \pi}{\sqrt{eE}L_0} \right|^3  \int_0^{\infty} \frac{{\rm d}\omega}{2\pi} \frac{\sqrt{eE}}{\omega} {\rm e}^{ 2 \frac{\omega}{\sqrt{eE}}\frac{\pi}{\sqrt{eE}L_0} } \left| \frac{\tilde{l}(\omega)}{L_0}\right|^2 \Biggl]. 
\end{align}
It is evident that the production number is suppressed exponentially by $L_0$ due to the overall factor $\exp[ -\pi (\pi/\sqrt{eE}L_0 )^2]$, i.e., the finite size effect strongly suppresses the Schwinger mechanism as well as the dynamical assistance.  Also, the third term (i.e., the dynamical assistance) is exponentially large $\propto \exp[ 2 (\omega/\sqrt{eE})(\pi/\sqrt{eE}L_0) ]$ compared to the first and second terms (i.e., the Schwinger mechanism with finite $L_0$).  This implies that the relative magnitude of the dynamical assistance increases with an exponential of $L_0^{-1}$.

For large $L_0$, the production number asymptotes the naive Schwinger formula with infinite $L_0 \to \infty$, i.e., no dynamical assistance, since $l_0/L_0 \to 0$.  Indeed, in the limit of $L_0 \to \infty$, Eq.~(\ref{eq-36}) behaves as
\begin{align}
	&\lim_{L_0 \to \infty} \overset{(-)}{N} \nonumber\\
	&= \frac{S_{\perp}L_0T}{(2\pi)^3} m^4 \times \left| \frac{eE}{m^2}  \right|^2 {\rm e}^{-\pi \frac{m^2}{eE}} \Biggl[
			1 + \frac{1}{L_0T} \tilde{l}(0) \nonumber\\
			&\quad
			+ \frac{1}{L_0T}\frac{1}{eE} \int_{-\infty}^{+\infty} {\rm d}p_z{\rm d}p'_z{\rm d}p_y \int_0^{\infty} {\rm d}\omega \;\left| \tilde{l}(\omega) \right|^2\nonumber\\
				&\quad\quad \times 
				 \frac{p_z^2}{eE} \frac{{p'_z}^2}{eE} {\rm e}^{- \pi \frac{p_y^2+\frac{p_z^2+{p_z'}^2}{2}}{eE} } \nonumber\\
				&\quad\quad\times \left| {}_1 \tilde{F}_1 \!\left(\! \frac{1}{2} \!+\! \frac{{\rm i}}{2}  \frac{m^2+p_y^2+p_z^2}{eE}; 1 \!+\! {\rm i} \frac{p_z^2-{p_z'}^2}{2eE} ; - {\rm i}\frac{\omega^2}{2eE} \!  \right) \! \right|^2 \!\Biggl], \label{e-q-50}
\end{align}
where we used
\begin{align}
	\frac{n\pi}{L_0} \to p_z,\ \frac{2\pi}{L_0}\sum_n \to \int_{-\infty}^{+\infty}{\rm d}p_z.  
\end{align}
Therefore, the dynamical assistance decays slowly $\propto L_0^{-1}$ for large $L_0$ if $l_0$ is fixed.  Note that, if one fixes the ratio $l_0/L_0$ instead of fixing $l_0$, the production number linearly increases with $L_0$.  This is simply because the dynamical assistance becomes stronger with $L_0$ since $l_0$ increases (i.e., the vibration becomes stronger) with $L_0$.

At intermediate $L_0$, the dynamical assistance overwhelms the finite size effect, and the production number becomes more abundant than the naive Schwinger formula for infinite $L_0 \to \infty$.

\subsubsection{$E$-dependence}\label{sec3b3}

\begin{figure}[!t]
\begin{center}
\includegraphics[clip, width=0.49\textwidth]{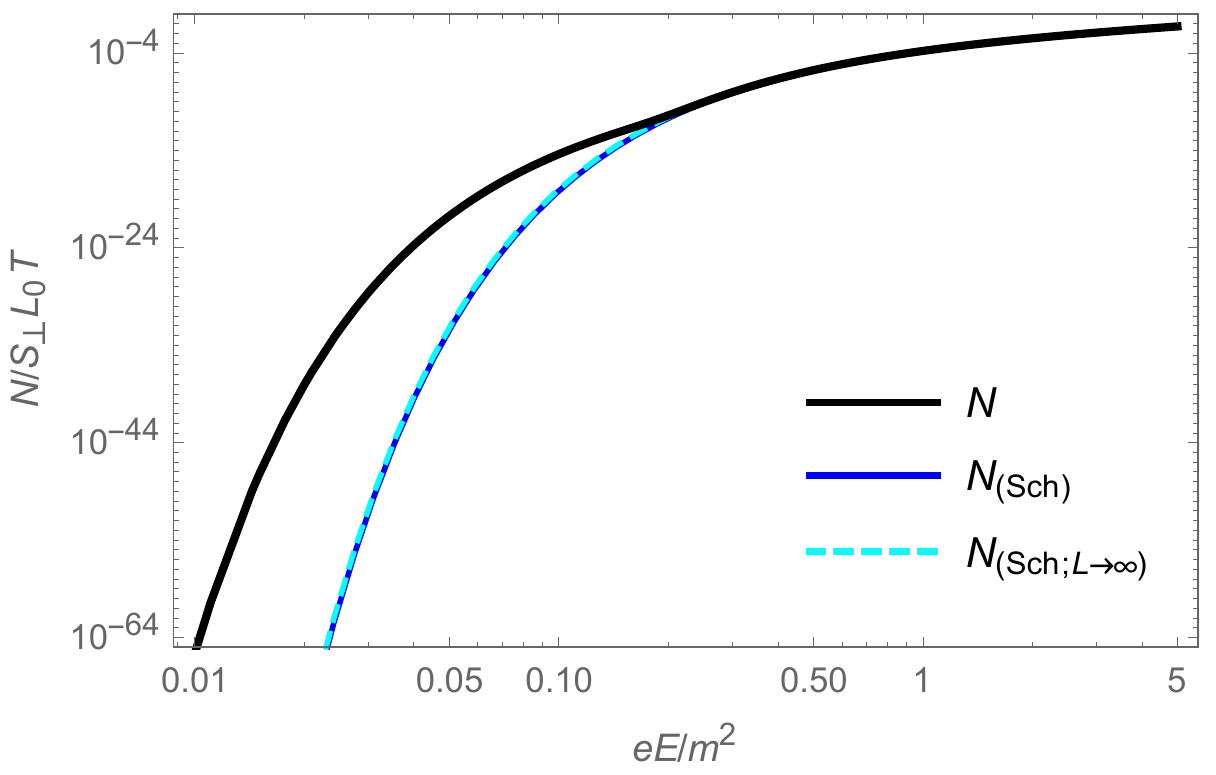}
\caption{\label{fig5} $E$-dependence of the total production number $N$ (\ref{eq-36}) for the monochromatic vibration (\ref{eq_-45}).  As a comparison, the production number for the Schwinger mechanism with finite $L$ (\ref{eq39}) and with infinite $L\to \infty$ (\ref{eq---42}) are shown in blue and dashed cyan, respectively.  The parameters are fixed as $\Omega/m=1, mL_0=30$, and $ml_0=0.3$.  }
\end{center}
\end{figure}

Figure~\ref{fig5} shows the electric-field strength $E$-dependence of the total production number $N$ (\ref{eq-36}).  As a demonstration, we considered sufficiently large system size $mL_0=30$, and a small vibration $ml_0=0.3$ (i.e., $l_0/L_0=1/100$).   We also fixed the frequency as $\Omega/m=1$, for which the naive dynamical Casimir effect expects no particle production $N_{\rm (Cas)}=0$ [Eq.~(\ref{eq--44})].

For a supercritical electric field exceeding the mass gap, the particle production is dominated by the Schwinger mechanism, and the dynamical assistance becomes unimportant.  This is because the Schwinger mechanism becomes free from the strong exponential suppression for a supercritical field, while the dynamical assistance is always suppressed by powers of the mass gap independent of the electric field strength.

For a subcritical electric field strength below the mass gap, the dynamical assistance becomes important, and the particle production is enhanced by orders of the magnitude compared to the naive Schwinger mechanism.  This is because the Schwinger mechanism is strongly suppressed by an exponential of the mass gap, while the dynamical assistance is suppressed just weakly by powers of the mass gap.

\section{Summary and discussion} \label{sec4}

We have discussed massive charged particle production from the vacuum in the presence of vibrating plates and a strong electric field.  Based on the perturbation theory in the Furry picture, we have analytically derived a formula for the particle production number, and have shown that (i) the formula smoothly describes the interplay between the Schwinger mechanism by the strong electric field and the dynamical Casimir effect by the vibrating plates, and the interplay is controlled by the typical frequency of the vibration; (ii) at intermediate frequency, the Schwinger mechanism and the dynamical Casimir effect assist with each other to dramatically enhance the particle production by orders of the magnitude; (iii) the dynamical assistance can be greater than the finite size effect, which gives an exponential suppression on the production number; and (iv) the dynamical assistance becomes more important for smaller system size and/or subcritical electric field strength.

Our results suggest a novel method to enhance the Schwinger mechanism by introduction of vibrating plates.  The vibration could be realized either mechanically or effectively, just as the usual experimental setups for the dynamical Casimir effect.  This could be interesting to the current intense laser experiments, in which the available field strength is still far below the critical strength $eE_{\rm cr}$ and hence some mechanism to enhance the Schwinger mechanism is highly demanded.  Our results indicate that the enhancement becomes larger for weaker electric field strength below the critical one, which is exactly the relevant parameter regime at the current experiments.  Note that one may combine the usual dynamically assisted Schwinger mechanism (i.e., superimposition of a fast electromagnetic field) \cite{sch08, piz09, dun09, mon10a, mon10b} to further enhance the production number.

One may interpret the above application in an opposite manner: One may use a strong electric field to enhance the dynamical Casimir effect.  Although the dynamical Casimir effect is usually applied for photon production, the same production mechanism, in principle, can be applied to massive particles (e.g., electron) as well.  However, it is very difficult within the current experimental technologies to achieve a very fast oscillation with frequency comparable to the mass scale of, e.g., electron, so that such massive particle production is not feasible in laboratory experiments at the present.  Our results suggest that one may lower the threshold frequency for the dynamical Casimir effect by applying a strong electric field if the massive particle is charged.  In other words, one may produce massive particles via the dynamical Casimir effect even with a slow vibration under the assistance by a strong electric field.

Another interesting application is ultra-relativistic heavy-ion collision experiments operated at Relativistic Heavy Ion Collider (RHIC) at Brookhaven National Laboratory and the Large Hadron Collider (LHC) at the European Organization for Nuclear Research (CERN).  In those experiments, expanding color flux tubes (sometimes called {\it glasma}) are produced just after a collision of ions \cite{low75, nus75, kov95a, kov95b, lap06}.  The speed of the expansion is relativistically fast, so that the dynamical Casimir effect may take place \cite{pro06} on top of the Schwinger mechanism by the chromo-electromagnetic field of the flux tubes.  If this is the case, the dynamical Casimir effect as well as the dynamical assistance between the two mechanisms could leave some experimental traces in, e.g., hadron multiplicities and rapidity correlations.

It is also interesting to pursue a condensed-matter analog of our results.  Electrical breakdown of materials (Landau-Zener transition \cite{lan32, zen32, stu32, maj32}) is one of the possible examples: Electrical breakdown can be understood as an analog of the Schwinger mechanism.  Usually, materials are assumed to be static in discussing electrical breakdown.  Our results suggest that addition of some mechanical perturbations onto materials (e.g., acoustic wave on the surface) and/or varying material properties (e.g., dielectric constant) in time may trigger electrical breakdown even below the naive threshold, which could be useful in designing, e.g., optical devices.

In the present paper, we have concentrated on the dynamical assistance between the Schwinger mechanism (electric field) and the dynamical Casimir effect (vibrating plates).  In principle, any kinds of external forces/fields can assist the Schwinger mechanism and vice versa.  In fact, no matter what the physics origin of the potential $V$ (\ref{eq16}) is, one gets the same assistance effect for the Schwinger mechanism for the same $V$.  It is, therefore, interesting to pursue other combinations of forces/fields.  For example, one may consider the assistance between the Schwinger mechanism and gravitational fields, which could be important to understand the magnetogenesis in the early Universe \cite{kob14, sob18, sha17, sor19, sta18, kob19}.  Another example is the axion production by axion condensate \cite{hua20}, which may take place at the early Universe where the curvature is large and hence the curvature effect could enhance the axion production and vice versa.  We leave these topics as future work.

\section*{Acknowledgments}

The author would like to thank Xu-Guang~Huang for collaboration at the early stage of this work, and Yukawa Institute for Theoretical Physics at Kyoto University (YITP) for their warm hospitality during a workshop ``Quantum kinetic theories in magnetic and vortical fields (YITP-T-19-06),'' where a part of this work was done.  The author was supported by National Natural Science Foundation in China (NSFC) under Grant No.~11847206.

\end{document}